\documentclass[useAMS,usenatbib]{mn2e} 
\usepackage{amsmath,amssymb}
\usepackage{graphicx,mathptm} 
\usepackage{times}
\usepackage{natbib}
\usepackage{psfig}

\def\kms{\mbox{km\,s$^{-1}$}}
\def\vvb{\bf v_B}
\def\vv{\bf v}
\def\hmpc{\ {\rm h^{-1}Mpc}}
\def\dd{{\rm d}}
\def\ltsim{\lower.5ex\hbox{$\; \buildrel < \over \sim \;$}}
\def\gtsim{\lower.5ex\hbox{$\; \buildrel > \over \sim \;$}}
\def\ltsim{\lower.5ex\hbox{$\; \buildrel < \over \sim \;$}}
\def\gtsim{\lower.5ex\hbox{$\; \buildrel > \over \sim \;$}}

\title[The velocity field of  2MRS galaxies]
{The velocity field of 2MRS, K$_{\rm s}$=11.75 galaxies: constraints on $\beta$ and  bulk flow from the luminosity function}
\author[Branchini et al.]  { Enzo Branchini$^{1,2,3}$, Marc Davis$^{4}$, Adi Nusser$^{5}$ \\
$^1$Dipartimento di Fisica ``E. Amaldi'', Universit\'a degli Studi ``Roma Tre'', via della Vasca Navale 84, 00146, Roma, Italy  \\
$^2$INFN Sezione di Roma Tre \\ 
$^3$INAF, Osservatorio Astronomico di Brera, Milano, Italy \\
$^4$Departments of Astronomy \& Physics, University of California, Berkeley, CA. 94720\\
$^5$Physics Department and the Asher Space Science Institute-Technion, Haifa 32000, Israel \\}

\begin{document}

\maketitle

\label{firstpage}

\begin{abstract}
Using the nearly full sky $K_s=11.75$ 2MASS Redshift Survey [2MRS]
of $\sim 45,000$ galaxies we reconstruct the 
underlying peculiar velocity field and
constrain the cosmological bulk flow within $\sim 100 \  \hmpc$. 
These results are obtained by maximizing the probability  to estimate the absolute
magnitude of a galaxy given its observed apparent magnitude and redshift.
At a depth of  $\approx 60 \hmpc$ we find a bulk flow
$\vvb$$=(90\pm 65, -230\pm 65, 50\pm 65) \ \kms$
in agreement with 
the theoretical predictions of the $\Lambda$CDM model.
The reconstructed peculiar velocity field $\vv$ that maximizes the likelihood
is characterized by the parameter $\beta=0.323 \pm 0.08$.
Both results are in agreement with those obtained previously using the $\sim 23,000$ galaxies 
of the shallower  $K_s=11.25$ 2MRS survey.

In our analysis we find that the luminosity function of 2MRS galaxies is poorly fitted by the Schechter form
and that luminosity evolves such that objects become fainter with increasing redshift 
according to $L(z)=L(z=0)(1+z)^{+2.7 \pm0.15}$. 

 \end{abstract}


\section{Introduction}

Peculiar velocities arising from the cosmological growth of density fluctuations affect the estimation
of the distances and luminosities of extragalactic objects from their measured redshifts, a spurious effect
commonly known as {\it redshift space distortions}. The effect is expected to be systematic because of the 
large scale coherence of the peculiar velocity field  in the $\Lambda$CDM model.
In this respect, redshift space distortions provide a unique tool to validate the $\Lambda$CDM and 
gravitational instability scenarios, to probe the underlying the velocity field and to 
estimate the growth rate of cosmological simulation $f(z)=\frac{d\ln{D}}{d\ln{a}}$, 
where $z$ is the redshift, $a=(1+z)^{-1}$ is the expansion parameter and $D$ is the linear growth factor. 
The growth rate $f$  mainly depends on the mass density parameter $\Omega$ and the redshift \citep{Peebles80}. 
The exact dependence is determined by  the underlying theory of gravity. 
Therefore the estimation of $f(z)$ constitutes a sensitive test to Einstein's General Relativity.

The availability of large and relatively deep galaxy redshift surveys in recent years has triggered 
a strong interest on the apparent anisotropy in galaxy clustering induced by redshift distortions since
they can be used
to tighten constraints over different cosmological parameters \citep{amendola2005} and  that to provide
a unique way to discriminate between a dark energy scenario and a modified gravity theory
\citep{guzzo2008, zhang2008}.  Redshift space distortions in the galaxy distribution at different 
epochs are now  regarded as one of the most effective ways to attack the dark energy problem
and constitutes one of the main scientific goals of
generation redshift surveys like BigBoss \citep{bigboss} or  Euclid \citep{laureijs2011}.

Redshift distortions not only misplace galaxies. They also affect the estimate of 
the objects' luminosities. And yet, little attention has been given to this second aspect 
of the same phenomenon.
In fact, the idea of using systematic biases in the estimated galaxy luminosities to constrain peculiar motions
is not new. It dates back to the work of \cite{TYS} who 
correlated the magnitudes of nearby galaxies with their redshifts to constrain the velocity of the Virgo 
cluster relative to the Local Group. 
However, the method requires a large number of objects to be effective. For this reason
several authors focused on average quantities rather than single objects. For example
\cite{baleisis1998, blake2002, Itoh} (and references therein) have exploited the Compton-Getting effect
and searched for dipole variation in the surface number density of distant galaxies 
to estimate the bulk flow. Similarly, \cite{abate11} looked for a dipolar modulation in the 
variation of a suitably defined average apparent magnitude across the sky. 
Currently available  all-sky redshift surveys, like 2MRS, 
allow an estimate of galaxy luminosities for a large number of galaxies.
In this case, peculiar motions can be inferred from  systematic variations in the 
estimated  galaxy luminosities across the sky.
\cite{NBD11} adopted this approach and showed that it can be used to 
estimate the the bulk flow in the local universe. 
c\cite{NBD12} took a step further and showed that the same idea could be used to constrain the growth rate of density 
fluctuations in the local universe, i.e. at $z\sim 0$.

In this paper we use the new redshift survey of  nearly  full-sky 2MASS Redshift Survey (2MRS)  of $\sim 45000$ galaxies with $K_s\le 11.75$ \citep{2mrs1175} to supersede the work of  \cite{NBD11} and \cite{NBD12} 
based on the previous release with a brighter apparent magnitude cut $K_s = 11.25$. 
The aim is twofold.
First of all, there is a considerable interest in large-scale flows with some controversial claims of  anomalous bulk flows 
on various scales that would exceed  $\Lambda$CDM prediction (see e.g. \cite{feldman010,Kash2012} and reference therein),
that were not confirmed by subsequent analyses (\cite{ND11,bilicki11,NBD11,turnbull012,osborne2011,mody2012}.
It is an intriguing issue that certainly justifies a closer look. Our technique provides a fresh approach to this
outstanding problem. Thanks to the improved dataset we should be able to detect significant departures from
$\Lambda$CDM predictions within $100 \ \hmpc$.
Second of all, our technique constrains the  velocity field independently on distance indicators.
As such, it is free of potential systematic errors arising from  a miscalibration of the
distance indicators.
From the spatial distribution and estimated luminosities of 2MRS galaxies we are able to model the 
linear velocity field and determine its only free parameter, $\beta \equiv f(\Omega)/b$, where $b$ is the linear  bias parameter
of the galaxy sample. Our aim is to improve the accuracy of the estimate obtained by  \cite{NBD12}
and constrain the fundamental quantity $f(\Omega)$ at $z\sim 0$.

Our method heavily relies on the estimation of the galaxy luminosity function [LF]. Therefore we take special care 
in detecting, evaluating and correcting for systematic biases related to the measurement of the LF.
For this purpose we use a suite of different methods, some of which completely new, 
to estimate  the LF and the selection functions of the sample. 
Since velocities are estimated at the redshifts of the objects, our results are probe to the so-called 
Kaiser rocket effect.  Any method aimed at estimating the underlying mass density field from a 
spatial distribution of mass tracers in a redshift survey should take this correction into account. 
Therefore, we  perform this correction and, in Appendix, we offer an analytic 
treatment of the effect.

The structure of the paper is as follows. In Section~\ref{sec:Theory} we describe the theoretical 
tools used in this work: the maximum likelihood method and the different LF estimators used in the 
analyses.
In Section~\ref{sec:data} we describe the real and simulated datasets.
In Section~\ref{sec:dnds} we introduce, implement and  apply a novel technique to estimate the 
selection function of the catalog directly from the observed redshift distribution of the galaxies.
The results are compared with those obtained in Section~\ref{sec:LF} in which we first compute the 
LF and by integration, we obtain the selection function.
In Sections~\ref{sec:bflow_data} and~\ref{sec:beta} we apply our maximum likelihood method to 
estimate the bulk flow and the $\beta$ parameter. Finally, in Section~\ref{sec:conc}
we discuss and summarize our main results.

\section{Theoretical tools}
\label{sec:Theory}

The scope of this work is to use the 2MRS flux limited at $K_s=11.75$
 to estimate the cosmological bulk flow 
 in the local universe and trace the underlying 
peculiar velocity field. Peculiar velocities contribute to the
measured redshift of an object.
If $r$ is the proper distance of a galaxy,
$z$ its redshift and  
$v$ is the line of sight component of its
peculiar velocity, then 
$s \equiv cz=r+v$, where all quantities are expressed in $\kms$, including the 
speed of light $c$.
The absolute magnitude is estimated from 
the apparent  magnitude, $m$, through 
$M_0=m-15-5\log{cz}$.
The "observed" magnitude is
different from the true one 
$M=m-15-5\log{r_l}=M_0-5\log{(1-v/cz)}$, where $r_l=r(1+z)$ is the luminosity distance.
The difference between $M$ and $M_0$ can 
be used to infer the peculiar velocity of the object. 
This can be done by  maximizing the probability $P(M_0|cz,\vv)$ of a galaxy having an observed magnitude $M_0$ given 
its redshift $cz$  and peculiar velocity $\vv$: 
\begin{equation}
P(M_0|cz,{\bf v})=\frac{\Phi(M)}{\int_{-\infty}^{M_l} \Phi(M) dM} \; ,
\label{eq:pmcz}
\end{equation}
where $\Phi(M)$ is the luminosity function, 
$M_{l}=m_l-25-5\log r_l$,
the apparent magnitude limit of the catalog is $m_l=K_s=11.75$
and the expression is valid as long as errors in the measured redshifts
are small ($\sigma_{cz}/cz \ll 1$).

\subsection{Bulk flow from magnitudes and redshifts}
\label{sec:bflow_tool}

\cite{NBD11} presented  a simple method to measure
cosmological bulk flows by minimizing 
systematic variations in the galaxy magnitudes estimated from the observed redshifts.

The method can be illustrated by the following example.
Let us assume that the peculiar velocity field is characterized by a bulk flow $\vvb$ and 
galaxies have a Schechter  LF  \citep{schechter}:
\begin{equation}
\label{eq:shformM}
\Phi(L)=0.4\ln(10) \Phi^* \left( \frac{L}{L_*}\right)^{1+\alpha}
{\rm exp}\left(-\frac{L}{L_*}\right)\; .
\end{equation}
In the large distance and small redshift error approximation
the probability in Eq.~\ref{eq:pmcz} is
 \begin{equation}
\label{eq:papp}
P(L_0|cz; {\bf v_B} )=\frac{0.4 \ln (10) \left(\frac{{\tilde L}_0}{L_*}\right)^{1+\alpha}{\rm e}^ {-{\tilde L}_0/L_*}}{\Gamma\left(1+\alpha,{\tilde L}_l/L_*\right)}\; ,
\end{equation}
where ${\tilde L}_0=(1-2 {\bf v_B}/cz)L_0$ and ${\tilde L}_l=(1-2{\bf v_B}/cz)L_l$
and absolute magnitudes are related to luminosities through $M=-2.5 \log L+const$.

The presence of a bulk flow $\vvb$ systematically increases or decreases
the estimated luminosity of a galaxy $L_0$. Therefore, an estimate of the bulk flow 
can be obtained by maximizing the probability $P(L_0|cz; {\bf v_B})$
with respect to $\vvb$. It is trivial to generalize this approach to a generic 
form of the luminosity function.

This method can be seen as a generalization of  the maximum likelihood approach 
proposed by  \cite{Itoh}. In that case the bulk flow was estimated from  
the apparent dipole anisotropy modulation in the surface number density 
of galaxies in the the SDSS-DR6 catalog, \citep{sdss6}.
To estimate the bulk flow that method requires angular positions,
apparent magnitudes and the  photometric redshift of the galaxies.
The method proposed here requires more information (spectroscopic redshifts)
but allows an estimate of the bulk flow from a differential quantity, the LF, 
rather than an integral one (the number density of objects).
As a result, this method  is more sensitive to bulk flows than 
the one proposed by  \cite{Itoh}. 
A similar method that does not use spectroscopic redshifts but only 
apparent magnitudes  has been recently proposed by \cite{abate11}.
In that case the authors looked for systematic, dipole-like variations 
in the apparent magnitude of the LRGs in the SDSS survey \citep{LRG}).

\subsection{$\beta$ from magnitudes and redshifts}
\label{sec:TheoryLFZ}

\cite{NBD12} have extended  the maximum likelihood technique discussed above
to constrain the full linear velocity field $\vv$.
The method requires redshifts, angular positions and apparent magnitudes of 
galaxies.
Galaxy positions are given in redshift space and are used
to compute the   linear velocity field as a function of  $\beta$.
Details of the computation of the linear velocity field can be found in \cite{ND94}.
Predicted velocities are used to compute distances and estimate the true absolute magnitude of the objects.
The best fit value of $\beta$  is found by maximizing the probability $P(M_0|cz,{\bf v}(\beta))$ over all galaxies in 
the sample.

We stress that the linear velocity field is predicted from the 
galaxy distribution in redshift space. Such a procedure is prone to the so-called 
Kaiser rocket effect, a systematic error induced by estimating the 
selection function of galaxies using  redshifts rather than 
distances. In this work we explicitly correct for this bias. A detailed 
treatment of the effect can be found in Appendix A.

\subsection{Estimators of the galaxy luminosity function}
\label{sec:LF}

The measurement of the LF
represents a key step in the maximum likelihood methods outlined above. 
To guarantee an accurate measurement and to minimize possible systematic errors
we have used different estimators for  the luminosity function, $\Phi(M)$, that we 
briefly describe below.

\begin{itemize}

\item $1/V_{\rm Max}$ estimator [$\Phi_V$]. This simple non-parametric estimator
originally proposed by \cite{schmidt}
weights each object by the maximum observable comoving volume in which 
it can be detected. It is the only estimator among those we have considered that is
 sensitive to large scale inhomogeneities in the galaxy distribution. 

\item  STY estimator  [$\Phi_{Sch}$]. 
This estimator has been originally proposed by \cite{STY}.
It  assumes a Schecther form  for $\Phi(M)$ \citep{schechter}
and computes the best fit parameters by maximizing
the product of the probabilities for galaxies to have 
a magnitude $M$ given the observed redshift $z$
(Eq.~\ref{eq:pmcz}):
\begin{equation}
P_s=\Pi_i P(M_{i}|z_i)\; .
\label{eq:psum}
\end{equation}
For a Schechter form 
\begin{eqnarray}
\label{eq:shform}
\nonumber
\Phi(M)&=&0.4\ln(10) \Phi^* 10^{0.4(\alpha+1)(M_*-M)}\\
&\times &{\rm exp}\left(-10^{0.4(M_*-M)}\right)\;
\end{eqnarray}
The free parameters  are $M_*$ and $\alpha$ 
and the normalization  $\Phi_*$. The latter does not concern us here.

\item Stepwise estimator  [$\Phi_{Stp}$]. 
This method, originally 
proposed by \cite{EEP}, is non-parametric and has been 
derived from a maximum likelihood approach.
The unknown luminosity function $\Phi(M)$
is discretized into $N_b-1 $ magnitude bins 
over the range
$M_1<M_2 \cdots <M_{N_b}$ so that
\begin{equation}
\Phi(M)=\Phi_i \quad {\rm for} \quad  M_{i+1}\ge  M>M_{i} \;  .
\end{equation} 
Given this stepwise form for $\Phi(M)$, the probability 
in Eq~\ref{eq:pmcz} becomes
\begin{equation}
P(M_i|z_i)=\frac{\Phi_i}{ (M_l-M_j) \Phi_j + \Delta M \sum_{k<j} \Phi_k}\; .
\end{equation}
where we take $M_{i+1}-M_i=\Delta M=const$ and assume
that the actual and limit magnitudes, $M$ and $M_l(z)$, fall into the bins $i$ and $j$, respectively. 
The $N_b$ free parameters $\Phi_i$ are estimated by maximizing the
product of single galaxies' probabilities.

\item Spline based estimator [$\Phi_{Spl}$].
The maximum likelihood method used to constrain the
velocity field $\vv$ requires a smooth LF in input.
For this reason we introduce a new estimator in which
the unknown LF is approximated by a smooth piecewise function
\begin{equation}
\Phi(M)=q_i(M) \quad {\rm for} \quad M_{i+1}\ge  M>M_{i} \; ,
\end{equation}
where $q_i$ is a third degree polynomial satisfying 
the boundary conditions $q_i(M_{i})= \Phi_i$ and $q_i(M_{i+1})= \Phi_{i+1}$
and defined such that its second derivative
$d^2\Phi/dM^2$ is continuous over the magnitude range $M_1-M_{N_b}$. 
The coefficients of the splines  can be efficiently computed using the standard techniques described in
\cite{NumRec}.
Splines can be integrated to estimate the denominator in Eq. ~\ref{eq:pmcz}
and the  coefficients $\Phi_i$ are determined by maximizing the 
product of probabilities $P_s=\Pi_jP_j(M_j|z_j)$ extended to all galaxies in the sample.
Splines obtained from this procedure may produce a noisy LF especially 
at the faint and bight end due to the limited number of objects. 
In order to suppress these spurious wiggles we maximize a function which is the sum of
$\log[P_s]$ and a penalty  function which acquires very large negative values
when the third derivative of the splines is large. 
This procedure efficiently suppresses the wiggles and still yields a set of best fit coefficients
$\Phi_i$ based on maximum likelihood considerations.
\end{itemize}


\section{Datasets}
\label{sec:data}

We use the recently compiled 2MRS catalog 
that contains  all galaxies brighter than $K_s=11.75$ with measured spectroscopic redshift
selected from the   2MASS XSC catalog of nearly one million objects  \citep{2mrs1175}. 
The catalog is  97.6 \% complete and mostly unaffected by interstellar extinction and stellar confusion
over the region $|b|>5^\circ $ for $30^\circ\le l \le 330^\circ$ and $|b|\ge 8^{\circ}$ otherwise. 
The total sky coverage is over  91 \%. 
The catalog contains about 43,000 galaxies and therefore
represents a significant improvement over the 
$K_s \le 11.25$ redshift catalog of $23,000$ objects used in the 
\cite{NBD11} and  \cite{NBD12} analyses.

Although the catalog extends out to $s \sim 30,000 \ \kms$ 
we restrict our analysis to objects at smaller distances. 
To minimize incompleteness for nearby objects
we consider a semi-volume limited sample 
that contains all galaxies with $s>s_{cut}$ and 
galaxies with $s\le s_{cut}$ that, if placed at $s_{cut}$, would be brighter 
than the magnitude limit. We set $s_{cut}=3,000 \ \kms$. 
The semi volume limited catalog is therefore obtained by excluding all galaxies  with 
$M>M_l(s_{cut})=m-5{\rm log} r_l(s_{cut}) -25$ and $s<s_{cut}$.
In addition, throughout  the paper  we work with a version of the survey 
with collapsed fingers-of-god in the main nearby clusters and 
with the masked region near the galactic plane
filled at a given redshift by folding the  the positions of galaxies  

In addition to the real catalog we will also consider a suite  of mock 2MRS catalogs.
They will be used to test the validity of our likelihood approach and assess its uncertainties.
Indeed we have used two different sets of mock catalogs:

\begin{itemize}

\item Mock catalogs used to test the bulk flow accuracy [$\vvb-$Mocks]:

These mock catalogs are the same as \cite{NBD11}.
The set is composed by 200 semi-volume limited mock catalogs 
with $s_{cut}=3,000 \ \kms$ which
contain the same number of objects as the real sample.
Mock galaxies  are randomly distributed  within a sphere of $200 \hmpc$.
Their absolute magnitudes are assigned according to the
LF  of 2MRS (early + late type) galaxies estimated by  \cite{w09}.
Redshifts of the objects are the sum of the Hubble flow and peculiar velocities modeled as
a random component sampled from a Gaussian distribution with zero mean and width of $300 \ \kms$.
This scatter accounts for  the combined effect of small scale velocity dispersion
and errors in the measured redshift.
No underlying large scale bulk flow was assigned to
the mock galaxies.

\item Mock catalogs used to test  the $\beta$ accuracy [$\beta-$Mocks]:

These mocks are the same ones used in \cite{NBD12}.
They are a set of  135 2MRS mock catalogs extracted  from the
mock Two Micron All Sky Survey  extracted from  the Millennium simulation.
Mock galaxies were obtained the semi analytic model of \cite{mill, delucia}.
In these mocks the central observer is not chosen at random. Instead, it
is selected to match the density and the dynamical properties of our
Local Group of galaxies. For our purposes the main relevant properties 
of the mock galaxies is their LF that is well approximated by a  Schechter form. 
More details on the mock 2MRS galaxies  can be found in  \cite{dn10}.

\end{itemize}

\section{Redshift distribution of 2MRS galaxies and their evolution}
\label{sec:dnds}

\begin{figure}
\includegraphics[width=0.49\textwidth]{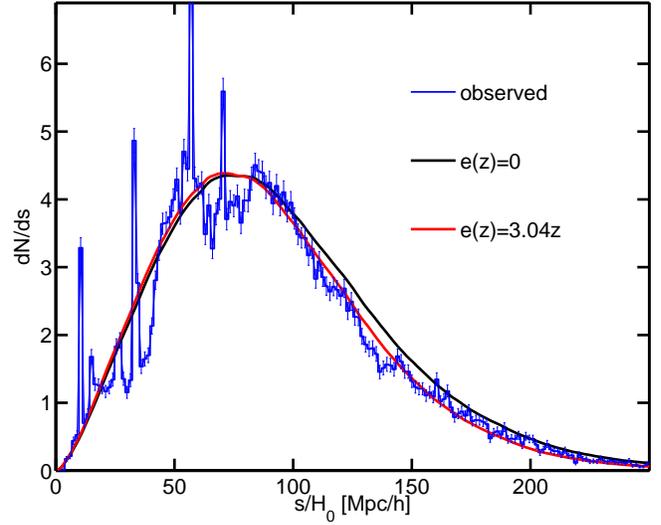}
\caption{
The blue histogram represents the observed redshift distribution of  2MRS galaxies
computed in redshift bins $\Delta s=150 \ \kms$. Errorbars represent 1 $\sigma$ Poisson uncertainties.
Continuos lines show the fits to the data obtained from the selection function computed using 
the {\it $F/T$ estimator}.
The black curve assumes \citet{Koch01}  k-correction but no luminosity evolution.
The red curve shows the effect of k-correction a luminosity  evolution $e(z)=3.04 z$.
}
\label{fig:DH}
\end{figure}

The blue histogram in figure \ref{fig:DH} shows 
the redshift distribution of objects in the
semi-volume limited catalog of 2MRS galaxies,  $dN/ds$.
We use constant redshift bin of size $\Delta s=150 \ \kms$. 
Errorbars represent the Poisson scatter in each bin 
$\sigma_N=\sqrt{N}=\sqrt{dN/ds \times \Delta s}$.
The black, continuous curve is a parametric fit to the distribution
\begin{equation}
\frac{dN}{ds}=A S(s) s^2 \; ,
\label{eq:dnds}
\end{equation}
where $S(s) $ is the galaxy selection function, i.e. the 
fraction of galaxies  in the catalog at 
redshift $s$.
The selection function can either be 
measured  directly from the observed counts \citep{KOS,DH82}
or estimated from the LF.  In this paper we adopt both approaches.
In this section we use the first one and compare the result
with the alternative approach in the next Section.
For this purpose we present a novel method, the  {\it $F/T$ estimator}, 
in which the selection function is computed 
by integrating the following equation
\begin{equation}
\frac{\dd \ln S(s)}{\dd s}  \Delta s =-\frac{F(s)}{T(s)}\; ,
\label{eq:FT}
\end{equation}
where  $T(s)$ is the number of galaxies  with redshift smaller than
$s$  that could also be detected at larger distances
while $F(s)$ is the number of galaxies within $s$ 
that can only be detected out to $s+\Delta s$.
The normalization $A$ is set by matching the  total number of 
galaxies in the catalog within $s=12,000 \ \kms$. 
This estimator assumes that the galaxy luminosity 
and selection functions do not depend on the environment.

In Fig.~\ref{fig:DH} we compare
the smooth redshift distribution of galaxies obtained from the selection function 
estimated with the  {\it $F/T$} method (black continuous curve) is compared with the 
observed $dN/ds$ (blue histogram).
The curve fits the data well out to $s=10,000 \  \kms$. At higher redshifts the 
{\it $F/T$ } method overestimates the observed number of galaxies. 
If real, this difference would indicate that we are located within an under-dense region
extending out to $\sim 20,000 \ \kms$.  
Alternatively, the mismatch can hint some systematic errors is affecting the data.
Before considering the first option,  let us consider the possible  sources of
systematic uncertainties. 
To estimate the ratio $F(s)/T(s)$ in Eq.~\ref{eq:FT} one needs to compute 
the rest-frame absolute magnitude of galaxies, i.e to apply the so-called 
 $k$-correction, $k(z)$. In addition, the galaxy luminosity may evolve with time according to some law, 
 $e(z)$.
We correct for these systematic effects in our magnitude estimation as follows:
\begin{equation}
M=m-25-5\log r_l(z)-k(z)-e(z)\; .
\label{eq:mags}
\end{equation}
In the $K_s$ band and at low redshift ($z<0.25$) the k-correction is negative.
 \footnote{Since the $K_s$ band is on the Rayleigh Jeans part of the spectrum, and assuming the galaxy has little warm
dust,  $k(z)$ is negative, unlike the optical bands.}
For 2MRS galaxies \cite{Koch01}, have found that $k(z)=-6\log(1+z)$, a corrections that we adopted in this work.
The evolution correction is more uncertain. It can be estimated by forcing 
a good match between the black curve and the histogram in Fig.~\ref{fig:DH}.
In practice we assume simple luminosity evolution model $L(z)=L(z=0)(1+z)^{\epsilon}$
and find $\epsilon$ by minimizing the $\chi^2$ function
\begin{equation}
\chi^2=\sum^{Nbins}_{i=1}\frac{1}{\sigma_{n,i}^2} \left[\frac{\dd N}{\dd s}-A S(s) s^2   \right]^2_{i}\; ,
\end{equation}
where  the summation is over all redshift bins and $\sigma_n$ is the Poisson noise in the galaxy counts.
We obtain $L(z)=L(z=0)(1+z)^{+2.7 \pm0.15}$.
The result of this correction is represented by the solid, red line in Fig.~\ref{fig:DH}, which indeed provides a 
good fit to observations.

\section{The luminosity function of 2MRS galaxies}
\label{sec:LF}

\begin{figure}
\includegraphics[width=0.49\textwidth]{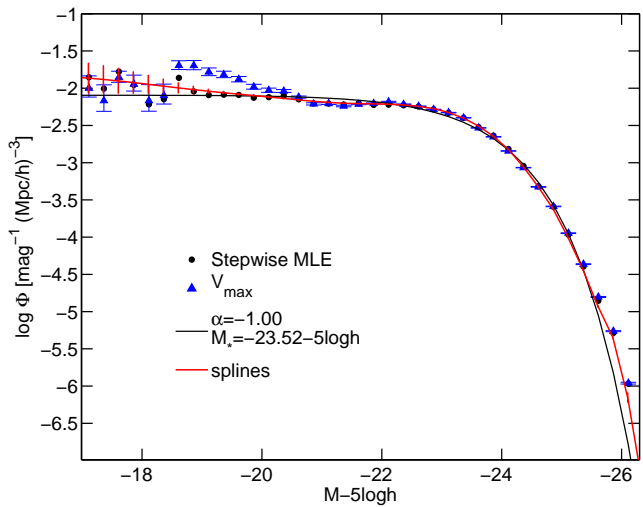}
\caption{
LFs of 2MRS galaxies obtained from the different estimators described in the text.
$\Phi_V$: blue triangles.  $\Phi_{Sch}$: Black, continuous curve.  $\Phi_{Stp}$: black dots.
$\Phi_{Spl}$: red, continuous line. Errorbars represent 1-$\sigma$ bootstrap errors. 
Luminosity functions are computed in bins  $0.25$ mag. Symbols are plotted
in every other bin for clarity.
}
\label{fig:phim}
\end{figure}

In this section we use the different estimators described in Section~\ref{sec:LF}
 to compute the LF of the  2MRS galaxies and compare the results to assess their 
 robustness.
 Absolute magnitudes were computed in redshift space using the
$k(z)$ and $e(z)$ corrections described above. LFs are
estimated in bins of 0.25 mag.  
The results are plotted in Fig.~\ref{fig:phim}. 
All LFs are normalized to match the  number of galaxies in the bin at $M=-23.37$ 
which is close to the value of $M_*=-25.52-5 \log h$ obtained from the Schechter fit. 

In the plot the blue triangles represent the LF obtained using the $1/V_{MAX}$ method
which is  sensitive to large scale inhomogeneities. There is an excess 
of faint ($M_{K_s}>-20 + 5 \log h$) objects  that we interpret as an
overdensity in the number of galaxies in the nearby ($s<2,200 \ \kms$) volume of the universe.
The black, thin curve shows the  Schechter LF with best fit parameters
$(\alpha=-1.0, M_*=-25.52-5 \log h)$. At the bright end this fit
deviates significantly from the others.
This is not surprising and reflects that fact that we are forcing 
a  Schechter fit to the LF of a composite sample of early and late type galaxies when
we know that the two subsamples are well fitted by two different 
Schechter forms \citep{w09} with $M_*$ for early type objects
i0.5 magnitude brighter than for the late-type objects \citep{NBD11}.
The results of the $\Phi_{Stp}$ (black dots) and $\Phi_{Spl}$ (continuous, red curve) estimators 
are consistent with each other and  with those of $\Phi_V$ 
for relatively bright objects $M_{K_s}-5 \log h<-20$, i.e. in a volume large enough for
fluctuations  in the number density of galaxies to be negligible.

Errorbars  in the plot are estimated from a bootstrap resampling analysis 
based on  20 random catalogs obtained from the original one by replacing each 
galaxy with $n_r$ objects,  where $n_r$ is drawn from a Poisson distribution function with mean unity.
All LF estimators have been applied to the 20 bootstrap catalogs.
The error bars represent the $rms$ scatter among the catalogs. Bootstrap uncertainties turned out to 
be very close  to Poisson errors.

LFs are estimated in redshift space. We do not expect that they are different from those measured 
in real space, i.e. placing each objects at their true distances.
The reason is that correction for peculiar velocities are of the order of $\langle (v/cz)^2\rangle $, where 
the average is taken over all directions in the surveyed area. Since we are dealing with an almost all-sky 
survey the effect is expected to be negligible.
To verify this hypothesis we have placed galaxies at their estimated true positions using the peculiar velocity model
discussed in  Section~\ref{sec:beta}. We  found no significant difference between the LFs estimated in real and redshift space.

\begin{figure}
\includegraphics[width=0.49\textwidth]{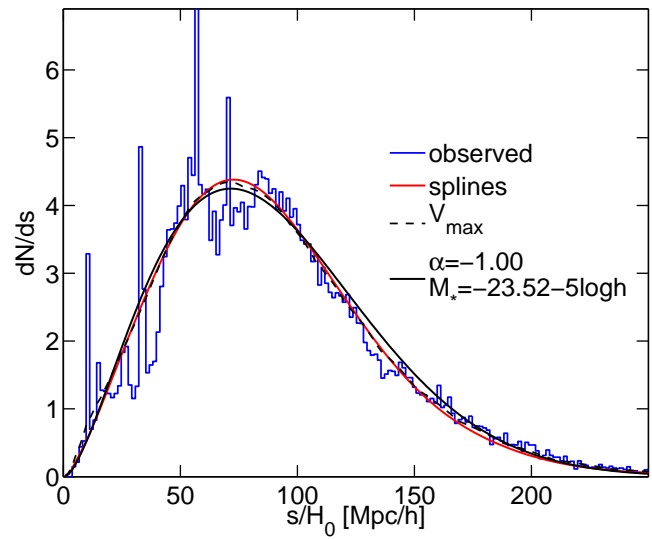}
\caption{
Redshift distribution of 2MRS galaxies. Blue histogram:  same $dN/ds$
as in Fig.~\ref{fig:DH}. Curves: $dN/ds$ predicted from different LF estimators.
Black, continuous : $\Phi_{Sch}$. Red, continuos:  $\Phi_{Spl}$.  Black, dashed $\Phi_{V}$. }
\label{fig:dndz}
\end{figure}

From the estimated LF it is straightforward to compute 
the galaxy selection function:
\begin{equation}
S(s)=\frac{\int_{-\infty}^{M_l(s)}\Phi(M)d M }{\int_{-\infty}^{M_l(s_{cut})}\Phi(M)d M}\; 
\label{eq:selfunct}
\end{equation}
for $s>s_{cut}$. In Fig.~\ref{fig:dndz}. we show the $dN/ds$ computed from
the estimated selection function using Eq.~\ref{eq:dnds}.
The different curves refer to different LF estimators. All of them are normalized to the 
observed galaxy counts (represented by the histogram) within $12,000 \ \kms$.

The  $dN/ds$ curves obtained from $\Phi_{Sch}$ (black continuous),  $\Phi_{Ste}$ (not plotted) and $\Phi_{Spl}$ (red continuous)
are independent of the underlying inhomogeneities in the galaxy distribution
unlike the one obtained from $\Phi_{V}$ (black, short-dashed)
This difference is explains why the $dN/ds$ obtained from the $1/V_{Max}$ estimator
is different from the other curves both in the nearby region and 
in correspondence of the peaks of the distribution.
The galaxy counts predicted from the Schecther LF 
are systematically larger above the observed one at  $s>10,000 \kms$. 
This discrepancy reflects the fact that the Schechter fit underestimates the number of bright and 
faint objects alike  (Fig.~\ref{fig:phim}), resulting in an artificially  small denominator in Eq.~\ref{eq:selfunct}. 

We notice  the remarkable similarity between 
the $dN/ds$ curves obtained from the spline and  $1/V_{Max}$ estimators 
and the one obtained from the  {\it $F/T$} method shown in figure \ref{fig:DH}. 
A result that further demonstrates the goodness of the new
{\it $F/T$ } estimator proposed in this paper.

\begin{figure}
\includegraphics[width=0.49\textwidth]{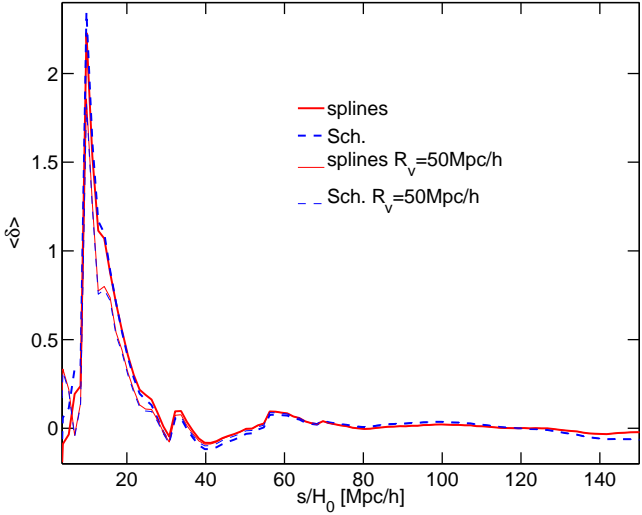}
\caption{
Mean galaxy overdensity  (Y-axis) within the redshift $s$ (X-axis) 
estimated from  $\Phi_{Sch}$ (blue, dashed curves) and  $\Phi_{Spl}$ (red, solid).
Thick and thin curves indicate the two redshift cuts of $3,000$ and $5,000 \ \kms$
applied to extracted the semi-volume limited sample.
All curves are normalized to $\langle \delta (s \le 12,000 \ \kms) \rangle=0$.  }
\label{fig:mean}
\end{figure}

From the estimated LFs it is straightforward to compute the mean galaxy overdensity within $s$:
\begin{equation}
1+ \langle \delta(<s) \rangle=\frac{\int_{-\infty}^{M_l(s)}\Phi(M)d M }{\int_{-\infty}^{M_l(s_{Norm})}\Phi(M)d M}\;.
\end{equation}
The mean is set by the number density of galaxies within the redshift $s_{Norm}=12,000  \ \kms$
chosen to normalize the selection function.
The estimated cumulative overdensity obtained from the different LF estimators is
 plotted in figure~\ref{fig:mean}.
For the sake of clarity we only show the overdensity computed from  the Schechter LF (thick, blue dashed line)
and from the spline estimator (think red)
since the latter is very similar to those obtained from
the $1/V_{Max}$ and the stepwise methods.
The overdensity predicted by the two estimators are very similar within $s =12,000 \ \kms$.
At larger distances, where the  Schechter fit underestimates the number of expected counts,
the corresponding overdensity is systematically smaller than obtained with the other estimators.
These results  depend on the choice of $s_{cut}$. Therefore 
we have repeated the  analysis using a m ore aggressive cut  $s_{cut}=5,000 \ \kms$. 
The results of are shown by the  thin curves in Fig.~\ref{fig:mean}. 
Changing $s_{cut}$ only affects the estimate of the overdensity
in the inner region ($s < 3,000 \ \kms$), as expected, but has no impact on the
results obtained at larger radii.

\section{Measuring the bulk flow from LF}
\label{sec:bflow_data}

We now apply the method described in Section \ref{sec:bflow_tool}
to the 2MRS $K_s <1.75$ galaxy catalog
to compute the bulk flow.
Unlike \cite{NBD11} we do not break down the sample into early and late type objects since
our best LF estimators $\Phi_{Spl}$ and  $\Phi_{Ste}$ are designed to 
deal with a mixed population of objects. We take the LF estimated from $\Phi_{Spl}$ as 
the reference case and will test the sensitivity of the results to the choice of the estimator.
We do not considered the Schechter LF since,
as we have shown, it underestimates the abundance of bright galaxies.

We estimate the bulk flow
in spherical shells of  depth $\Delta s=4,000 \ \kms$ and out to 
$s=10,000 \ \kms$ by maximizing the probability function
$P_s=\Pi_i P_i(M_0|cz,\vvb )$, where the product is over all galaxies
and $P_i$ is the probability of the single object 
defined in Eq.~\ref{eq:pmzc}. 
The  binning in redshift and the volume sampled are the same as in
 \cite{NBD11} but the number of objects is twice as large.
The results 
are displayed in Fig.~\ref{fig:B_on_shell}.
Points connected by lines represent the  Cartesian components of the differential bulk flow
estimated at the mean radius of each redshift shell. Different line styles are used for the 
different Cartesian components specified by labels.
Errorbars show the {\it rms}  scatter in the 
bulk flows estimated from the 200 $\vvb-$Mocks.
These errors are contributed by shot noise, uncertainties in the observed magnitude and redshifts.
They do not include cosmic variance. 
We find that the error budget is dominated by shot noise rather than redshift uncertainties..
Errors in the measured magnitudes induce the same systematic 
errors in both the measured and the reference LFs and do not affect the 
bulk flow estimate.
Not surprisingly, the results are remarkably robust to the method used to
estimate the luminosity function: 
$\Phi_{Spl}$ and  $\Phi_{Ste}$  and $\Phi_V$. 
Finally, results are also robust to the 
 $k$-correction and/or the  evolution corrections adopted, as we have verified by 
 switching off either corrections and  yet obtaining very similar bulk flows.

Our results can be compared directly with those of \citep{NBD11} and 
shown in Fig.1 of their paper.
The two bulk flows are fully consistent with each other at all redshifts and 
for all Cartesian components.
For example, at  $R\approx 6,000 \ \kms$ we find a bulk flow of 
$\vvb$ $ =(90\pm 65, -230\pm 65, 50\pm 65) \ \kms$, fully consistent with 
the one of \citep{NBD11}: $\vvb $$=(100\pm 90, -240\pm 90, 0\pm 90)\kms$.
The fact that errors scale with the square root of the objects  confirms that
errors are dominated by Poisson noise.

\begin{figure}
\includegraphics[width=0.49\textwidth]{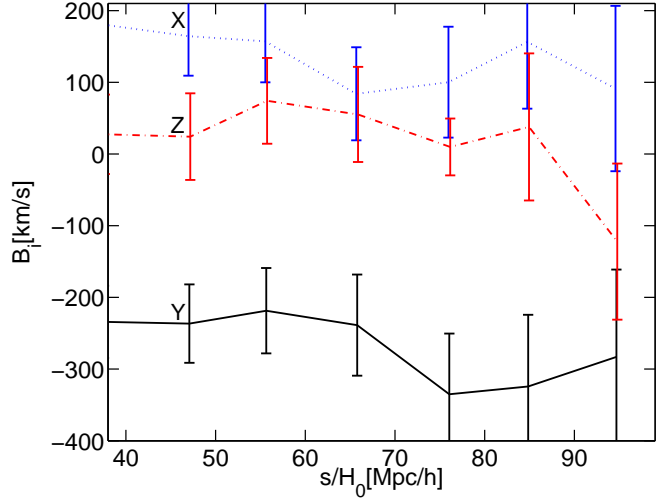}
\caption{
The bulk flow of 2MRS galaxies with $K_s<12.75$ estimated from their luminosities in 
spherical shells $4,000 \ \kms$-thick. The three Cartesian components are represented by
a dotted (X), solid (Y) and dot-dashed (Z) lines.}
\label{fig:bulk}
\end{figure}

\section{Measuring $\beta$ from LF}
\label{sec:beta}

In this section we apply the method described in Section~\ref{sec:TheoryLFZ}
to model the linear velocity field traced by the 2MRS galaxies and to
estimate the parameter $\beta$. The method 
uses the gravity field computed from the spatial distribution
of the galaxies in the survey to minimize the scatter between the 
observed and the expected luminosity function. The result provides
an estimate of $\beta$. The gravity field is computed from the 
spatial distribution of galaxies in redshift space
using the linear method  of \cite{ND94}.
As a reference luminosity function we use the one determined with the 
spline estimator  assuming the luminosity evolution and $k$-correction
discussed in Section~\ref{sec:dnds}.
Since the velocity field is found by maximizing $P_s=\Pi_j P_j(M_0|z,\vv(\beta)) $
with respect to $\beta$ we need to produce a suite of model 
velocity fields for different values of $\beta$. This is done by 
running the  \cite{ND94} reconstruction procedure using 
different values of  $\beta$ in the range [0.1,1] in steps of $0.02$.

The method  is prone to several systematic errors:
\begin{itemize}
\item Since galaxy positions are given in redshift space, 
a direct estimate of the mass overdensity would systematically 
affect the predicted peculiar velocities. We correct for 
this so-called Kaiser rocket effect as described in Appendix A.
\item   To apply linear theory we need to filter out nonlinear contributions. 
For this purpose we first smooth the 
galaxy distribution  with a Gaussian window 
of radius  $400 \ \kms$ and then apply a second Gaussian filter of radius 
$R_s$ to remove residual nonlinearities. We choose $R_s=600 \ \kms$
as a reference case and evaluate the robustness of the results to the choice of $R_s$.
\item  Linear theory  recovers the flow pattern  reasonably  well 
up to   $\delta \ltsim 1$, \citep{nussetal91,BEN02}.  For this reason
 \cite{NBD12} removed all galaxies in regions with overdensity above
 $\delta_{cut}=1$ and verified that the results did not change significantly 
 when a more conservative threshold $\delta_{cut}=2$ was adopted. 
 In our analysis we do not use any threshold and test the robustness of the 
 results to overdensity cuts. 
 
\end{itemize}

\begin{figure}
\includegraphics[width=0.49\textwidth]{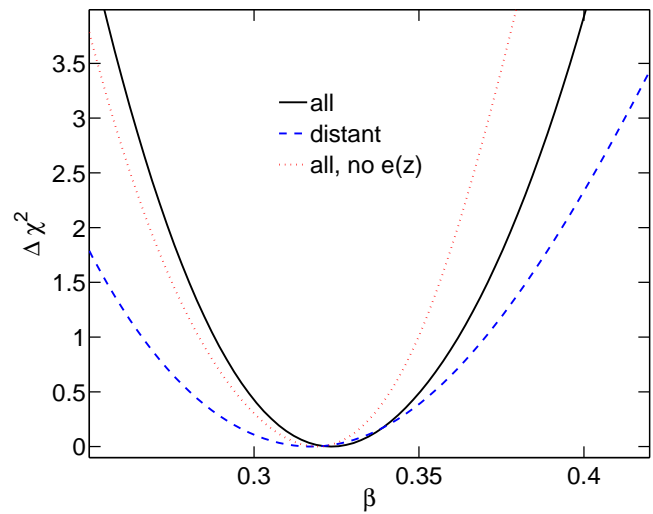}
\caption{
$\Delta \chi^2$ as a function of $\beta$. Solid line refers to all galaxies within 
$10,000 \ \kms$ with magnitudes corrected 
for luminosity evolution. The best fit value is $\beta=0.323 \pm 0.035$, where the quoted error
is the width of the curve at $\Delta \chi^2=1$ and 
account for the finite number of galaxies. 
Red dashed curve: same galaxy sample but no  correction for luminosity evolution. ($\beta= 0.319 \pm 0.033$).
Blue dashed: subsample with galaxies at $s>4,000 \ \kms$.  ($\beta =0.317 \pm  0.05$).  }
\label{fig:chibeta}
\end{figure}

Given the model velocity field $\vv(\beta)$ and the estimated galaxy luminosity function
 we apply the  \cite{ND94}  method to all  galaxies within $10,000 \ \kms$ and minimize 
$-\ln P_s=  -\sum_i \ln P(M_{0i}|cz_i; v(\beta))$ with respect to $\beta$.
We do not consider objects beyond  $10,000 \ \kms$
due to the rapid decline in the observed  number density of galaxies.
in Fig.~\ref{fig:chibeta} we show the quantity $\Delta \chi^2=-2\ln P_s(\beta) +2\ln P_s(\beta_{min}) $,
where  $\beta_{min}$ is the best fit value of $\beta$ found at the minimum of the curve.
The width of the curve at $\Delta \chi^2=1$ provides an estimate of the 1-$\sigma$ error. When
all objects are considered (black, continuous curve) we find  $ \beta =0.323 \pm 0.035$.
to be compared with $ \beta =0.35 \pm 0.05$ found by \cite{NBD12}. The errors quoted here 
are mainly contributed by shot noise, as confirmed by the  scaling with the number of galaxies.
However, other sources contribute to the error budget. 
We obtain more realistic errors by repeating the analysis
on the 135 $\beta-$Mock catalogs. These errors  
include contributions from cosmic variance, nonlinear effects as well as shot noise.
We find that cosmic variance and shot noise are similar in size and that the best 
fit estimate is $ \beta =0.323 \pm 0.083$.
 
To detect possible systematic errors and check the robustness of the results we have repeated the likelihood procedure
changing the free parameters of the model.
\begin{itemize}
\item The method relies on the measurement of the galaxy LF.
We have checked that the choice of the
estimator has a negligible effect on the $\beta$ value. 
This result is not surprising since, as shown
explicitly in  \cite{NBD12} the form of the   luminosity function only  affects the weighting given to galaxies in a  certain luminosity range but has no impact on the value of the best fit $\beta$. 
\item 
The choice of galaxy evolution model adopted $e(z)$ might also affect the results.
However, the effect is expected to be small since we only consider objects within $s=10,000 \ \kms$ where
the effect of the evolution is small, as shown in Fig.~\ref{fig:DH}. 
To verify this assumption we have repeated the analysis by switching off luminosity evolution, $e(z)=0$. The results, shown by
by the red dotted curve  in Fig.~\ref{fig:chibeta}, confirm that the the effect
is negligible
\item To assess the impact of nonlinearities associated to the peaks of the density field
we have repeated the analysis excluding all galaxies above  $\delta_{cut}=1$ and  $\delta_{cut}=2$,
as in \cite{NBD12}. We find no significant change in the best fit $\beta$ value.
\item The smoothing radius $R_s$ may also affect the estimate of $\beta$.
Increasing the smoothing radius decreases the amplitude of the 
gravity field so that a larger value of $\beta$ is required to fit the data. We have estimated the 
amplitude of the effect by increasing the Gaussian smoothing  to $R_s= 1,000 \  \kms$. The 
best fit value of $\beta$ increases, as expected. However, the amplitude of this shift is $\sim 3$ 
times smaller  than the random error.
\item The catalog incompleteness for faint objects preferentially affects the innermost part of the sample. To minimize 
possible systematics errors we have repeated the analysis excluding all objects 
within $s=4000 \ \kms$. The result is illustrated by the blue, dashed curve in Fig.~\ref{fig:chibeta}.
The  main effect is to increase shot-noise errors (by $\sim 40 $ \%)
with a negligible shift in the best fit  $\beta$. 
\end{itemize}

\section{Discussion and Conclusions}
\label{sec:conc}

In this work we have used the new 2MRS galaxy catalog of galaxies  brighter  than $K_s=11.75$ to 
estimate the peculiar velocity field and its bulk flow in the nearby ($s \le 10,000  \ \kms$) universe
from the apparent brightening/dimming of galaxy luminosities. To do this 
we have used the same maximum likelihood 
techniques proposed by  \cite{NBD11} and  \cite{NBD12} and applied to the brighter
2MRS $K_s=11.25$ sample, which contains about $\sim 50 \%$ less objects.
Since our technique heavily relies on the knowledge of the luminosity and selection 
functions of the galaxies, most of our efforts have been devoted in measuring
accurately these quantities to keep possible systematic errors under control. 

The main results of our analysis can be summarized as follows:
\begin{itemize}
\item We have computed the LF of 2MRS galaxies down to 
$M_{K_s}-5\log{h}=-17$ using four different estimators to 
search for possible biases in the measured LF (see e.g.~\cite{ilbert04}) and derived quantities.
The LF obtained from the $1/V_{Max}$ estimator shows an excess of 
objects fainter than $M_{K_s}-5\log{h}=-20$ not seen with other estimators.
We interpret this feature as large scale overdensity in the distribution
of nearby ($cz<2,200 \ \kms$) galaxies since the $1/V_{Max}$ method 
 is sensitive to large scale inhomogeneities.

Parametric estimators in which the LF is modeled with a  Schechter form may
induce systematic errors. More specifically, we find that the best fit Schecther form
to the observed LF systematically underestimates  the number density of bright 
($M_{K_s}-5\log{h}<25.5$) objects. This bias arises from 
fitting a single Schechter luminosity function to a composite sample of early and late
type objects  \citep{NBD11}. 
We have introduced a novel, non-parametric estimator for the LF similar to the 
stepwise method of  \cite{EEP} with the advantage of producing a smooth 
LF to be fed into the maximum likelihood method used in this work.
Both the new estimator and the stepwise method hint at an excess of faint objects,
quantified by a steepening of the LF in excess of the typical power-law slope $\alpha=-1.0$.
The significance of this faint end excess, however, is barely above 1 $\sigma$.

It turns out that the differences among the LFs estimated with various method are 
rather small and have no practical impact for our goals. That is to say that the 
bulk flow and the $\beta$ values obtained from the maximum likelihood procedure
do not significantly depend on LF estimator adopted.

\item From the LFs we estimate the selection function of the catalog and 
the redshift distribution of the 2MRS galaxies. 
The results are compared with the observed $dN/ds$ and with the redshift distribution 
obtained from a novel {\it F/T} estimator. 
This new statistical tools allows one to measure the selection function of the catalog from 
the observed redshift distribution of the galaxies, rather than from the LF.
The {\it F/T} estimator systematically over predicts the number of objects in the outer 
part of the survey that we interpret as an evolution in the galaxy luminosity.
According to this interpretation the galaxy luminosity evolves so that objects 
would grow fainter with the redshift. A simple luminosity evolution model $L(z)=L(z=0)(1+z)^{+2.7 \pm0.15}$
is sufficient to fit to the observed $dN/ds$.

This result is not surprising since galaxy evolution in the infrared band can be significant even in the local patch of the universe.
In the far infra red, evidence for an evolution in the number density of
IRAS galaxies has been reported by a number of authors, although there is some controversy on the 
magnitude of the effect ( see e.g. \cite{oliver92,Fisher95,springel98,takeuchi03} and reference therein).
However, all these works find that  the galaxy number density increase with redshift
whereas in our case the positive evolution in luminosity would decrease the number density
of objects selected above a given flux.
This puzzling result, however, does not affect the outcome of our maximum likelihood methods
since in our analysis  we consider galaxies with $s<10,000 \ \kms$, where the evolution is found to be negligible.

\item
Our estimation of the bulk flow in the local  ($s<10,000 \ \kms$) universe
is fully consistent with that  of \cite{NBD11}. 
For example, at  $R\approx 6,000 \ \kms$ we find
$\vvb$ $ =(90\pm 65, -230\pm 65, 50\pm 65) \ \kms$ to be compared with 
 $\vvb $$=(100\pm 90, -240\pm 90, 0\pm 90)\kms$.
In both cases errors are dominated by shot noise and which explains the 
$\sim 40 \%$ improvement in the accuracy of the estimate. Cosmic variance is not included.
This the most precise estimate of the bulk flow in the local universe
obtained without using distant indicators. 
Among alternative bulk flow estimators, 
the one used here is arguably more precise than those proposed by \cite{Itoh}
and \cite{abate11} since, in our case, we look for systematic variations in a differential
quantity, the LF, rather than using integral quantities. Our method is also superior to the one
proposed by  \cite{HT96} that exploits the kinetic Sunyaev-Zel'dovich CMB distortions along the line-of-sight
to galaxy clusters  \citep{ksz},
due to the limited number of available galaxy clusters within  $10,000 \ \kms$.
Indeed, the likelihood method proposed here is the idea tool to probe bulk flows 
at different locations in  the universe from luminosity variations measured in 
deep and wide next generations redshift surveys (for a quantitative assessment see \cite{NBD11}).

Our estimate of the bulk flow agrees  well with the recent estimates of the bulk flow 
of  \cite{ND11}  $\vvb$$(R\approx 6,000 \ \kms)=(120\pm 40, -250\pm 40, 40\pm 60)\kms$.
based on the SFI++ Tully-Fisher catalog of distance indicators \citep{mas06, spring07, dn10},
We also agree with the bulk flow obtained from SNe Ia data \citep{colin011,dai011} and most notably
with \cite{turnbull012}  who used the high quality 'First amendment' dataset.
Finally, our result  also agrees with the recent analysis of the 2MASS galaxy dipole \cite{bilicki11}.
It is quite remarkable that consistent bulk flows have been obtained from different datasets and
using different methods  affected by different systematics and that all of them are in agreement 
with $\Lambda$CDM expectations.

\item
Using the gravity field computed from the spatial distribution of 2MRS galaxies 
we were able to obtain a linear model for the velocity field in the local universe
by minimizing the scatter in the LF. From this procedure we obtain
$\beta=0.323 \pm 0.083$, in agreement with the results of \cite{NBD12} ($\beta=0.323\pm0.1$).
The increase in the accuracy is more modest than for the bulk flow since in this case the error budget 
is almost equally contributed by shot noise and cosmic variance, which is now accounted for.
We did check that this result is robust to luminosity evolution, 
to the choice of the LF estimator, to catalog incompleteness for faint objects and to the 
smoothing scale used to remove nonlinearities.
Our estimation of $\beta$ is also in good agreement with that of \cite{dn10} obtained by comparing the gravity field 
of 2MRS $K_s<11.25$ galaxies with the peculiar velocities in the SFI++ galaxy catalog
($\beta=0.325\pm0.045$) and also with that of \cite{bilicki11} ($\beta=0.38\pm0.04$)  
obtained from the 2MASS galaxy dipole.

From  $\beta$ it is possible to constrain  the growth rate of density fluctuations $f(\Omega)=\beta/b $
if $b$ can be determined independently.  However measuring the galaxy bias from existing datasets is difficult
and no theory of galaxy evolution is currently able to precisely constrain its value.
Alternatively, one could estimate some parameter combination that does not depend on bias and 
yet provides the possibility of efficiently discriminating  among different cosmological scenarios.
The combinations  $f(\Omega) \sigma_8=\beta\sigma_{8,gal}$ provides 
such possibility since both $\beta$ and $\sigma_{8,gal}$, the $rms$ density contrast  in galaxy number,
can be measured from redshift surveys and $f(\Omega) \sigma_8$ is as good as $f(\Omega)$
to test  alternative cosmological scenarios  \citep{White09,percival2009,song2009}.
The validity of this probe relies on the hypothesis that at 8 $\hmpc$, the scale at which one 
measures $\sigma_{8,gal}$, galaxy bias has the same value as at
the much larger scales in which one measures $\beta$.
Estimates of $f(\Omega) \sigma_8$ from the apparent  anisotropy in galaxy clustering 
have been obtained in the redshift range $z\approx[0.2,0.8]$ \citep{percival04,samushia12,blake11,guzzo2008}.
Several authors have pointed out that an accurate estimate of the growth rate at $z\approx 0$,
which can only be obtained from the peculiar velocity field in the local universe,  would significantly
increase the discriminatory power of this cosmological probe. 
The most recent estimates have been obtained by
\cite{dn10} ($f(\Omega) \sigma_8=0.31\pm0.06$ at $z<0.033$) and 
\cite{turnbull012}  ($f(\Omega) \sigma_8=0.4\pm0.04$ at $z\approx 0.02$). 
In our case, taking $\sigma_{8,gal}=0.97\pm0.05$ from  \cite{w09,reid2010}, 
we find $f(\Omega) \sigma_8=0.31\pm0.09$  for  $z<0.033$, in good agreement  with 
the other results.

Since in many models of modified gravity  the growth rate can be parametrized as
$f(\Omega)=\Omega^{\gamma}$ \citep{linder05}  many recent works have focused on 
comparing the estimated growth index $\gamma$ with the canonical value $6/11$ of a
$\Lambda$CDM universe. The value of  $\gamma$ from our estimate of  $f(\Omega) \sigma_8$
must be consistent with those obtained from the estimates of  \cite{dn10} \cite{turnbull012}.
The value obtained combining the two estimates ($\gamma=0.616\pm0.052$
Hudson and Turnbull 2012, {\it ApJ submitted}) is consistent with the one obtained by \cite{ND11} 
from the bulk flow   of SFI++ galaxies ($\gamma=0.495\pm0.096$) and with those obtained from 
galaxy clustering on larger scales.
All of them agree with a $\Lambda$CDM standard gravity model.

\end{itemize}
 
\section{Acknowledgments}
This work was supported by THE ISRAEL SCIENCE FOUNDATION (grant No.203/09), the German-Israeli Foundation for 
Research and Development,  the Asher Space Research
Institute and  by the  WINNIPEG  RESEARCH FUND.
MD acknowledges the support provided by the NSF grant  AST-0807630. 
EB  acknowledges the support provided by
MIUR PRIN 2008 'Dark energy and cosmology with large galaxy surveys' and by
Agenzia Spaziale Italiana (ASI-Uni Bologna-Astronomy Dept. 'Euclid-NIS' I/039/10/0)
AN thanks the Physics Department of the Roma Tre University
for the kind hospitality.

\appendix
\section{The Kaiser rocket effect}

In this appendix we work out the expression for the so-called Kaiser rocket effect for two different types
of catalogs: a flux limited  and a volume limited redshift survey.

\subsection{Kaiser effect for a flux limited survey}
Assume we have a flux limited redshift survey of galaxies with apparent magnitudes $<m_l$.
The selection function $S(r)$ is proportional to the average density and, apart from a normalization factor, is defined as
\begin{equation}
S(r)\propto \int_{-\infty}^{M_l(r)}\Phi(M)d M
\label{eq:selfu}
\end{equation}
where $M_l=m_l-15\log(r_l)-15$, $r_l$ is the luminosity distance in $\kms$ 
and $\Phi(M)$ is the galaxy luminosity function. 
Here, for convenience,  we drop the subscript $_l$ from the luminosity distance.
The mean number density of galaxies within $r_n$ is defined as
\begin{equation}
\bar n=\frac{3}{4\pi r_n^3}\sum_{{r_i < r_n}} \frac{1}{S(r_i)}\; .
\end{equation}
I assume that $S(r_n)$ is sufficiently large  to avoid using the other  estimator which 
has $J_3$ in it (that estimator minimizes the variance at the expense of biasing $\bar n$). 
The true density contrast in a cell of size $\Delta V$ at $r$ is then 
\begin{equation}
\label{eq:dt}
1+\delta_t=\frac{1}{\bar n}  \sum_{i\in \Delta V} \frac{1}{S(r_i)}=\frac{n_o}{\bar n} \frac{1}{S(r)}\; .
\end{equation}
where $n_o \Delta V$ is the actual number of  observed galaxies within $\Delta V$ and we have assumed that  $\Delta V $ is so small that $r_i=r$.  

Since we are given the redshift $s=r+u$ rather than distances, we can only compute 
$S$ at the redshift position $s_i$ of a galaxy. 
This will introduce a bias known as Kaiser rocket effect since it would induce a spurious
component in the gravitational attraction estimated in redshift space.
Note also that $\Delta V$ will change by the transformation to 
redshift space, which will amount to the usual redshift distortions.  
As a result, the estimated density contrast in redshift space  is
 in \ref{eq:dt},
\begin{equation}
1+\delta=\frac{n_o}{\bar n} \frac{1}{S(r)}=\frac{n_o}{\bar n} \frac{1}{S(s)}\left(1+\frac{1}{s} \frac{d\ln S}{d\ln s} u\right) =1+\delta_t\left(1+\frac{1}{s} \frac{d\ln S}{d\ln s} u\right)\; .
\end{equation}
where $\frac{n_o}{\bar n} \frac{1}{S(s)}$ are directly determined from the data. 
The extra term  involving the logarithmic derivative of $S$ is the Kaiser term and can
be evaluated at $s$ instead of $r$ when one assumes linear theory. 
Any method aimed at estimating the underlying mass density field from a 
spatial distribution of mass tracers in a redshift survey should take this correction into account. 

\subsection{Kaiser effect for a volume limited survey}

It is often assumed that Kaiser correction is negligible in a volume limited survey. We show here that this is not quite the case.
Le us consider a volume limited survey of all galaxies that are bright enough to be seen out 
to a distance $r_{max}$, i.e. all these galaxies have $M<M_{l,v}=m_l-5\log(r_{max})-15$. 
The true density contrast  at $r<r_{max} $ is 
\begin{equation}
1+\delta_t=\frac{n_o(r)}{\bar n}\; ,
\end{equation}
where ${\bar n}$ is defined as $3N/(4\pi r_{max}^3)$ where $N $ is the total number of galaxies in the volume limited  survey. 
However, since we observe redshifts and not distances  $r=s-u$, objects are selected according to 
their absolute magnitude estimated in redshift space
$M_0=m-5\log(s)-15\approx-5log(r)-2.17(u/s)-15$ and
$M=m-15\log(r)-15\approx M_0+2.17(u/s)$. 
Thus, for example, in a volume element $\Delta V$ where $u$ is positive, the physical limit  $M_{l,v}$ 
is larger than its redshift-space estimate  $M_{0,l,v}$. 
Therefore, for redshift space data, the mean number density within $s$ is
\begin{equation}
 n_o = \int_{-\infty}^{M_{lv}+2.17u/s}\Phi(M)dM\; .
\end{equation}
where the integration extends out to $M_{lv}+2.17(u/s)$
instead of $M_{lv}$.
To correct for this effect the density contrast should be defined as 
\begin{eqnarray}
1+\delta &=&\frac{n_o}{\bar n} \frac{ \int_{-\infty}^{M_{lv}}\Phi(M)dM }{\int_{-\infty}^{M_{lv}+2.17u/s}\Phi(M)dM}\\
&=& \frac{n_o}{\bar n} \frac{S(r_{max})}{S(r_{max}-u)}\\
&\approx&  \frac{n_o}{\bar n}\left[1+\left(\frac{1}{s} \frac{d\ln S}{d \ln r}\right)_{r_{max}}\frac{u}{r_{max}}\right]
\end{eqnarray}
where we have used the definition of  $S$ in
Eq.~\ref{eq:selfu}. 
This is the analogous of the Kaiser correction for  a flux limited survey except that 
$u$ is multiplied by a quantity that is defined at $r_{max}$ rather than at $r$. 
For the 2MRS $K_s<11.75$ catalog, $d\ln S/d\ln r$ grows from  $0.92 $ at $s=3,000 \ \kms$ to $1.42$ at $5,000 \ \kms$ and the corresponding Kaiser term in a sub-volume limited catalog cannot be 
neglected.

\bibliographystyle{mn2e} 
\bibliography{LFest_9}

\begin{thebibliography}{}

\bibitem[\protect\citeauthoryear{{Abate} \& {Feldman}}{{Abate} \&
  {Feldman}}{2011}]{abate11}
{Abate} A.,  {Feldman} H.~A.,  2011, ArXiv e-prints

\bibitem[\protect\citeauthoryear{{Adelman-McCarthy}, {Ag{\"u}eros}, {Allam},
  {Allende Prieto}, {Anderson} \& {Anderson}}{{Adelman-McCarthy}
  et~al.}{2008}]{sdss6}
{Adelman-McCarthy} J.~K.,  {Ag{\"u}eros} M.~A.,  {Allam} S.~S.,  {Allende
  Prieto} C.,  {Anderson} K.~S.~J.,    {Anderson} S.~F. e.~a.,  2008, \apjs,
  175, 297

\bibitem[\protect\citeauthoryear{{Amendola}, {Quercellini} \&
  {Giallongo}}{{Amendola} et~al.}{2005}]{amendola2005}
{Amendola} L.,  {Quercellini} C.,    {Giallongo} E.,  2005, \mnras, 357, 429

\bibitem[\protect\citeauthoryear{{Baleisis}, {Lahav}, {Loan} \&
  {Wall}}{{Baleisis} et~al.}{1998}]{baleisis1998}
{Baleisis} A.,  {Lahav} O.,  {Loan} A.~J.,    {Wall} J.~V.,  1998, \mnras, 297,
  545

\bibitem[\protect\citeauthoryear{{Bilicki}, {Chodorowski}, {Jarrett} \&
  {Mamon}}{{Bilicki} et~al.}{2011}]{bilicki11}
{Bilicki} M.,  {Chodorowski} M.,  {Jarrett} T.,    {Mamon} G.~A.,  2011, \apj,
  741, 31

\bibitem[\protect\citeauthoryear{{Blake}, {Brough}, {Colless}, {Contreras},
  {Couch}, {Croom}, {Davis} \& {Drinkwater}}{{Blake} et~al.}{2011}]{blake11}
{Blake} C.,  {Brough} S.,  {Colless} M.,  {Contreras} C.,  {Couch} W.,  {Croom}
  S.,  {Davis} T.,    {Drinkwater} M.~J.,  2011, \mnras, 415, 2876

\bibitem[\protect\citeauthoryear{{Blake} \& {Wall}}{{Blake} \&
  {Wall}}{2002}]{blake2002}
{Blake} C.,  {Wall} J.,  2002, \nat, 416, 150

\bibitem[\protect\citeauthoryear{{Branchini}, {Eldar} \& {Nusser}}{{Branchini}
  et~al.}{2002}]{BEN02}
{Branchini} E.,  {Eldar} A.,    {Nusser} A.,  2002, \mnras, 335, 53

\bibitem[\protect\citeauthoryear{{Colin}, {Mohayaee}, {Sarkar} \&
  {Shafieloo}}{{Colin} et~al.}{2011}]{colin011}
{Colin} J.,  {Mohayaee} R.,  {Sarkar} S.,    {Shafieloo} A.,  2011, \mnras,
  414, 264

\bibitem[\protect\citeauthoryear{{Dai}, {Kinney} \& {Stojkovic}}{{Dai}
  et~al.}{2011}]{dai011}
{Dai} D.-C.,  {Kinney} W.~H.,    {Stojkovic} D.,  2011, \jcap, 4, 15

\bibitem[\protect\citeauthoryear{{Davis} \& {Huchra}}{{Davis} \&
  {Huchra}}{1982}]{DH82}
{Davis} M.,  {Huchra} J.,  1982, \apj, 254, 437

\bibitem[\protect\citeauthoryear{{Davis}, {Nusser}, {Masters}, {Springob},
  {Huchra} \& {Lemson}}{{Davis} et~al.}{2011}]{dn10}
{Davis} M.,  {Nusser} A.,  {Masters} K.~L.,  {Springob} C.,  {Huchra} J.~P.,
  {Lemson} G.,  2011, \mnras, 413, 2906

\bibitem[\protect\citeauthoryear{{De Lucia} \& {Blaizot}}{{De Lucia} \&
  {Blaizot}}{2007}]{delucia}
{De Lucia} G.,  {Blaizot} J.,  2007, \mnras, 375, 2

\bibitem[\protect\citeauthoryear{{Efstathiou}, {Ellis} \&
  {Peterson}}{{Efstathiou} et~al.}{1988}]{EEP}
{Efstathiou} G.,  {Ellis} R.~S.,    {Peterson} B.~A.,  1988, \mnras, 232, 431

\bibitem[\protect\citeauthoryear{{Eisenstein}, {Annis}, {Gunn}, {Szalay},
  {Connolly}, {Nichol} \& {Bahcall}}{{Eisenstein} et~al.}{2001}]{LRG}
{Eisenstein} D.~J.,  {Annis} J.,  {Gunn} J.~E.,  {Szalay} A.~S.,  {Connolly}
  A.~J.,  {Nichol} R.~C.,    {Bahcall} e.~a.,  2001, \aj, 122, 2267

\bibitem[\protect\citeauthoryear{{Feldman}, {Watkins} \& {Hudson}}{{Feldman}
  et~al.}{2010}]{feldman010}
{Feldman} H.~A.,  {Watkins} R.,    {Hudson} M.~J.,  2010, \mnras, 407, 2328

\bibitem[\protect\citeauthoryear{{Fisher}, {Huchra}, {Strauss}, {Davis},
  {Yahil} \& {Schlegel}}{{Fisher} et~al.}{1995}]{Fisher95}
{Fisher} K.~B.,  {Huchra} J.~P.,  {Strauss} M.~A.,  {Davis} M.,  {Yahil} A.,
  {Schlegel} D.,  1995, \apjs, 100, 69

\bibitem[\protect\citeauthoryear{{Guzzo} et~al.,}{{Guzzo}
  et~al.}{2008}]{guzzo2008}
{Guzzo} L.,  et~al., 2008, \nat, 451, 541

\bibitem[\protect\citeauthoryear{{Haehnelt} \& {Tegmark}}{{Haehnelt} \&
  {Tegmark}}{1996}]{HT96}
{Haehnelt} M.~G.,  {Tegmark} M.,  1996, \mnras, 279, 545

\bibitem[\protect\citeauthoryear{{Huchra}, {Macri}, {Masters}, {Jarrett},
  {Berlind}, {Calkins} \& {Crook}}{{Huchra} et~al.}{2011}]{2mrs1175}
{Huchra} J.~P.,  {Macri} L.~M.,  {Masters} K.~L.,  {Jarrett} T.~H.,  {Berlind}
  P.,  {Calkins} M.,    {Crook} A.~C. e.~a.,  2011, ArXiv e-prints

\bibitem[\protect\citeauthoryear{{Ilbert}, {Tresse}, {Arnouts}, {Zucca},
  {Bardelli}, {Zamorani}, {Adami}, {Cappi}, {Garilli}, {Le F{\`e}vre},
  {Maccagni}, {Meneux}, {Scaramella}, {Scodeggio}, {Vettolani} \&
  {Zanichelli}}{{Ilbert} et~al.}{2004}]{ilbert04}
{Ilbert} O.,  {Tresse} L.,  {Arnouts} S.,  {Zucca} E.,  {Bardelli} S.,
  {Zamorani} G.,  {Adami} C.,  {Cappi} A.,  {Garilli} B.,  {Le F{\`e}vre} O.,
  {Maccagni} D.,  {Meneux} B.,  {Scaramella} R.,  {Scodeggio} M.,  {Vettolani}
  G.,    {Zanichelli} A.,  2004, \mnras, 351, 541

\bibitem[\protect\citeauthoryear{{Itoh}, {Yahata} \& {Takada}}{{Itoh}
  et~al.}{2010}]{Itoh}
{Itoh} Y.,  {Yahata} K.,    {Takada} M.,  2010, \prd, 82, 043530

\bibitem[\protect\citeauthoryear{{Kashlinsky}, {Atrio-Barandela} \&
  {Ebeling}}{{Kashlinsky} et~al.}{2012}]{Kash2012}
{Kashlinsky} A.,  {Atrio-Barandela} F.,    {Ebeling} H.,  2012, ArXiv e-prints

\bibitem[\protect\citeauthoryear{{Kirshner}, {Oemler} Jr. \&
  {Schechter}}{{Kirshner} et~al.}{1979}]{KOS}
{Kirshner} R.~P.,  {Oemler} Jr. A.,    {Schechter} P.~L.,  1979, \aj, 84, 951

\bibitem[\protect\citeauthoryear{{Kochanek}, {Pahre}, {Falco}, {Huchra},
  {Mader}, {Jarrett}, {Chester}, {Cutri} \& {Schneider}}{{Kochanek}
  et~al.}{2001}]{Koch01}
{Kochanek} C.~S.,  {Pahre} M.~A.,  {Falco} E.~E.,  {Huchra} J.~P.,  {Mader} J.,
   {Jarrett} T.~H.,  {Chester} T.,  {Cutri} R.,    {Schneider} S.~E.,  2001,
  \apj, 560, 566

\bibitem[\protect\citeauthoryear{{Laureijs}, {Amiaux}, {Arduini},
  {Augu{\`e}res}, {Brinchmann}, {Cole}, {Cropper}, {Dabin}, {Duvet} \&
  {Ealet}}{{Laureijs} et~al.}{2011}]{laureijs2011}
{Laureijs} R.,  {Amiaux} J.,  {Arduini} S.,  {Augu{\`e}res} J.~.,  {Brinchmann}
  J.,  {Cole} R.,  {Cropper} M.,  {Dabin} C.,  {Duvet} L.,    {Ealet} A. e.~a.,
   2011, ArXiv e-prints

\bibitem[\protect\citeauthoryear{{Linder}}{{Linder}}{2005}]{linder05}
{Linder} E.~V.,  2005, \prd, 72, 043529

\bibitem[\protect\citeauthoryear{{Masters}, {Springob}, {Haynes} \&
  {Giovanelli}}{{Masters} et~al.}{2006}]{mas06}
{Masters} K.~L.,  {Springob} C.~M.,  {Haynes} M.~P.,    {Giovanelli} R.,  2006,
  \apj, 653, 861

\bibitem[\protect\citeauthoryear{{Mody} \& {Hajian}}{{Mody} \&
  {Hajian}}{2012}]{mody2012}
{Mody} K.,  {Hajian} A.,  2012, ArXiv e-prints

\bibitem[\protect\citeauthoryear{{Nusser}, {Branchini} \& {Davis}}{{Nusser}
  et~al.}{2011}]{NBD11}
{Nusser} A.,  {Branchini} E.,    {Davis} M.,  2011, \apj, 735, 77

\bibitem[\protect\citeauthoryear{{Nusser}, {Branchini} \& {Davis}}{{Nusser}
  et~al.}{2012}]{NBD12}
{Nusser} A.,  {Branchini} E.,    {Davis} M.,  2012, \apj, 744, 193

\bibitem[\protect\citeauthoryear{{Nusser} \& {Davis}}{{Nusser} \&
  {Davis}}{1994}]{ND94}
{Nusser} A.,  {Davis} M.,  1994, \apjl, 421, L1

\bibitem[\protect\citeauthoryear{{Nusser} \& {Davis}}{{Nusser} \&
  {Davis}}{2011}]{ND11}
{Nusser} A.,  {Davis} M.,  2011, \apj, 736, 93

\bibitem[\protect\citeauthoryear{{Nusser}, {Dekel}, {Bertschinger} \&
  {Blumenthal}}{{Nusser} et~al.}{1991}]{nussetal91}
{Nusser} A.,  {Dekel} A.,  {Bertschinger} E.,    {Blumenthal} G.~R.,  1991,
  \apj, 379, 6

\bibitem[\protect\citeauthoryear{{Oliver}, {Rowan-Robinson} \&
  {Saunders}}{{Oliver} et~al.}{1992}]{oliver92}
{Oliver} S.~J.,  {Rowan-Robinson} M.,    {Saunders} W.,  1992, \mnras, 256, 15P

\bibitem[\protect\citeauthoryear{{Osborne}, {Mak}, {Church} \&
  {Pierpaoli}}{{Osborne} et~al.}{2011}]{osborne2011}
{Osborne} S.~J.,  {Mak} D.~S.~Y.,  {Church} S.~E.,    {Pierpaoli} E.,  2011,
  \apj, 737, 98

\bibitem[\protect\citeauthoryear{{Peebles}}{{Peebles}}{1980}]{Peebles80}
{Peebles} P.~J.~E.,  1980, {The large-scale structure of the universe}

\bibitem[\protect\citeauthoryear{{Percival}, {Burkey}, {Heavens}, {Taylor},
  {Cole}, {Peacock}, {Baugh} \& {Bland-Hawthorn}}{{Percival}
  et~al.}{2004}]{percival04}
{Percival} W.~J.,  {Burkey} D.,  {Heavens} A.,  {Taylor} A.,  {Cole} S.,
  {Peacock} J.~A.,  {Baugh} C.~M.,    {Bland-Hawthorn} J.,  2004, \mnras, 353,
  1201

\bibitem[\protect\citeauthoryear{{Percival} \& {White}}{{Percival} \&
  {White}}{2009}]{percival2009}
{Percival} W.~J.,  {White} M.,  2009, \mnras, 393, 297

\bibitem[\protect\citeauthoryear{{Press}, {Teukolsky}, {Vetterling} \&
  {Flannery}}{{Press} et~al.}{1992}]{NumRec}
{Press} W.~H.,  {Teukolsky} S.~A.,  {Vetterling} W.~T.,    {Flannery} B.~P.,
  1992, {Numerical recipes in FORTRAN. The art of scientific computing}

\bibitem[\protect\citeauthoryear{{Reid}, {Percival}, {Eisenstein}, {Verde},
  {Spergel}, {Skibba}, {Bahcall}, {Budavari}, {Frieman}, {Fukugita} \&
  {Gott}}{{Reid} et~al.}{2010}]{reid2010}
{Reid} B.~A.,  {Percival} W.~J.,  {Eisenstein} D.~J.,  {Verde} L.,  {Spergel}
  D.~N.,  {Skibba} R.~A.,  {Bahcall} N.~A.,  {Budavari} T.,  {Frieman} J.~A.,
  {Fukugita} M.,    {Gott} J.~R.,  2010, \mnras, 404, 60

\bibitem[\protect\citeauthoryear{{Samushia}, {Percival} \&
  {Raccanelli}}{{Samushia} et~al.}{2012}]{samushia12}
{Samushia} L.,  {Percival} W.~J.,    {Raccanelli} A.,  2012, \mnras, 420, 2102

\bibitem[\protect\citeauthoryear{{Sandage}, {Tammann} \& {Yahil}}{{Sandage}
  et~al.}{1979}]{STY}
{Sandage} A.,  {Tammann} G.~A.,    {Yahil} A.,  1979, \apj, 232, 352

\bibitem[\protect\citeauthoryear{{Schechter}}{{Schechter}}{1980}]{schechter}
{Schechter} P.~L.,  1980, \aj, 85, 801

\bibitem[\protect\citeauthoryear{{Schlegel}, {Abdalla}, {Abraham} \&
  {Ahn}}{{Schlegel} et~al.}{2011}]{bigboss}
{Schlegel} D.,  {Abdalla} F.,  {Abraham} T.,    {Ahn} C. e.~a.,  2011, ArXiv
  e-prints

\bibitem[\protect\citeauthoryear{{Schmidt}}{{Schmidt}}{1968}]{schmidt}
{Schmidt} M.,  1968, \apj, 151, 393

\bibitem[\protect\citeauthoryear{{Song} \& {Percival}}{{Song} \&
  {Percival}}{2009}]{song2009}
{Song} Y.,  {Percival} W.~J.,  2009, \jcap, 10, 4

\bibitem[\protect\citeauthoryear{{Springel} \& {White}}{{Springel} \&
  {White}}{1998}]{springel98}
{Springel} V.,  {White} S.~D.~M.,  1998, \mnras, 298, 143

\bibitem[\protect\citeauthoryear{{Springel}, {White}, {Jenkins}, {Frenk},
  {Yoshida}, {Gao}, {Navarro}, {Thacker}, {Croton}, {Helly}, {Peacock}, {Cole},
  {Thomas}, {Couchman}, {Evrard}, {Colberg} \& {Pearce}}{{Springel}
  et~al.}{2005}]{mill}
{Springel} V.,  {White} S.~D.~M.,  {Jenkins} A.,  {Frenk} C.~S.,  {Yoshida} N.,
   {Gao} L.,  {Navarro} J.,  {Thacker} R.,  {Croton} D.,  {Helly} J.,
  {Peacock} J.~A.,  {Cole} S.,  {Thomas} P.,  {Couchman} H.,  {Evrard} A.,
  {Colberg} J.,    {Pearce} F.,  2005, \nat, 435, 629

\bibitem[\protect\citeauthoryear{{Springob}, {Masters}, {Haynes}, {Giovanelli}
  \& {Marinoni}}{{Springob} et~al.}{2007}]{spring07}
{Springob} C.~M.,  {Masters} K.~L.,  {Haynes} M.~P.,  {Giovanelli} R.,
  {Marinoni} C.,  2007, \apjs, 172, 599

\bibitem[\protect\citeauthoryear{{Sunyaev} \& {Zeldovich}}{{Sunyaev} \&
  {Zeldovich}}{1980}]{ksz}
{Sunyaev} R.~A.,  {Zeldovich} I.~B.,  1980, \mnras, 190, 413

\bibitem[\protect\citeauthoryear{{Takeuchi}, {Yoshikawa} \& {Ishii}}{{Takeuchi}
  et~al.}{2003}]{takeuchi03}
{Takeuchi} T.~T.,  {Yoshikawa} K.,    {Ishii} T.~T.,  2003, \apjl, 587, L89

\bibitem[\protect\citeauthoryear{{Tammann}, {Yahil} \& {Sandage}}{{Tammann}
  et~al.}{1979}]{TYS}
{Tammann} G.~A.,  {Yahil} A.,    {Sandage} A.,  1979, \apj, 234, 775

\bibitem[\protect\citeauthoryear{{Turnbull}, {Hudson}, {Feldman}, {Hicken},
  {Kirshner} \& {Watkins}}{{Turnbull} et~al.}{2012}]{turnbull012}
{Turnbull} S.~J.,  {Hudson} M.~J.,  {Feldman} H.~A.,  {Hicken} M.,  {Kirshner}
  R.~P.,    {Watkins} R.,  2012, \mnras, 420, 447

\bibitem[\protect\citeauthoryear{{Westover}}{{Westover}}{2007}]{w09}
{Westover} M.,  2007, PhD thesis, Harvard University

\bibitem[\protect\citeauthoryear{White, Song \& Percival}{White
  et~al.}{2009}]{White09}
White M.,  Song Y.-S.,    Percival W.~J.,  2009, Mon. Not. Roy. Astron. Soc.,
  397, 1348

\bibitem[\protect\citeauthoryear{{Zhang}, {Yu}, {Noh} \& {Zhu}}{{Zhang}
  et~al.}{2008}]{zhang2008}
{Zhang} H.,  {Yu} H.,  {Noh} H.,    {Zhu} Z.-H.,  2008, Physics Letters B, 665,
  319

\end{thebibliography}

\label{lastpage}
\label{lastpage}

\end{document}